# Comparison of stable spin textures in in-plane vs. out-of-plane magnetized exchange-biased multilayers

B.Dieny[1*], O.Fruchart[1], E.E. Marinero[2]

*corresponding author: bernard.dieny@cea.fr

1. Univ. Grenoble Alpes, CEA, CNRS, SPINTEC, 38000 Grenoble, France
2. School of Materials Engineering, Neil Armstrong Hall of Engineering, Purdue University, 701 W Stadium Ave, West Lafayette, IN 47907, United States

## Abstract

This paper delves into the origins and specificity of the unique stable spin textures (360° closed loop domain walls and skyrmions) observed in exchange-biased systems, with either in-plane or out-of-plane magnetic anisotropy. In the case of skyrmions, which are nanometer-sized bubbles consisting of closed-loop 180° walls in perpendicularly-magnetized media, the stability of these spin textures arises from the existence of Dzyaloshinskii-Moriya Interactions (DMI). These interactions induce chirality of the domain walls, yielding to some extent a so-called topological protection. More complex structures such as skyrmoniums have been observed, consisting of closed loop 360° walls. Recently, skyrmions formed in the absence of an applied external magnetic field have been stabilized in exchange biased out-of-plane magnetized systems. About two decades ago, another type of stable spin-textures were observed in exchange biased systems, with in-plane magnetization, in particular in the pinned reference layer of spin-valves. These textures consist of 360°-domain-wall rings, the stability of which arises from the easy-plane shape anisotropy of these layers. In this paper, we compare these spin-textures and highlight the similarities and differences in their formation, structure and origin of their stability.

1.  Introduction:

The shift of the hysteresis loop of a ferromagnet (FM) when grown adjacent to an antiferromagnet (AFM), **the exchange bias effect**, was discovered by Meiklejohn and Bean in 1957 [1]. It arises from interfacial magnetic exchange coupling between the constituent magnetic layers. These thin-film FM/AFM bilayers are model systems to investigate and understand the underlying physics of exchange bias. The unidirectional pinning of the ferromagnet resulting from exchange bias can be established in a FM/AFM bilayer by cooling it in an external magnetic field from above the AFM blocking temperature or by applying a magnetic field during thin film growth [1]. As a result, the hysteresis loop of the FM layer exhibits a field-shift away from the origin, whose magnitude is called the *exchange bias field*. This effect has been of scientific interest for decades, and it depends on the magnetic properties and the microstructure of the FM and AFM materials. It is also exploited in a wide variety of magnetic devices to pin the magnetization of the reference layer in metallic spin-valves or in-plane magnetized magnetic tunnel junctions (MTJ) used in particular as hard-disk-drive (HDD) readers [2], as well as magnetic field sensors [3], amongst others. Besides, the electrical manipulation of exchange bias opens new possibilities for the development of low-energy consumption spintronic devices [4].

Numerous studies have investigated the origins and engineering of exchange bias in FM/AFM structures. Salient findings from these studies include: exchange coupling between the ferromagnet and the antiferromagnet is long-ranged in nature (it has been observed across 5.5 nm-thick nonmagnetic spacers) [5]; during magnetization reversal, the AFM spins may form exchange-spring domain wall structures extending along the thickness on both sides across the FM/AFM interface; the external field was found to move only the ferromagnetic part of the hybrid domain walls, while the antiferromagnetic parts of the hybrid domain walls defining the AFM domains remain unaffected [6]; 360° domain-wall rings may form upon field-driven magnetization reversal of in-plane-magnetized films, resulting from spatial fluctuations of the exchange-bias field [7]. These rings constitute a type of magnetic texture topologically protected by its close loop shape and remaining stable up to fields of several tens of mT.

Although exchange bias has been mainly studied and used in spin-valves and tunnel junctions magnetized in-plane, it is not limited to such systems. The existence of perpendicular exchange biasing for FM/AFM multilayers in Co/Pt multilayers exchanged coupled to CoO thin layers was demonstrated by Maat *et al.* [8]. The interfacial exchange energies were found to be twice larger for in-plane vs. out-of-plane field cooling geometry. This was attributed to differences in spin direction in the AFM after cooling. Subsequently, perpendicular exchange bias was reported with FeMn [9] as well as in synthetic (Co/Pt)/Ru/(Co/Pt)/FeMn exchange biased ferrimagnetic pinned layer [10].

Exchange bias was also observed in systems consisting of layers with orthogonal anisotropies. Marinero and coworkers studied the magnetic exchange interactions in Fe/TbFe wherein the Fe layer has in-plane magnetization (Fe) while the ferromagnetic TbFe exhibits out-of-plane anisotropy [11, 12]. The interfacial coupling was shown to result in canting of the TbFe spins away from the perpendicular easy axis (perpendicular), and be responsible for the observed exchange bias effect [13]. Orthogonal exchange bias was also reported in NiFe magnetized in-plane and exchange-coupled to out-of-plane magnetized (Pt/Co) multilayers [14,15]. In these systems, exchange bias appears when the (Pt/Co) multilayer is in a multidomain configuration and results from interfacial exchange interaction between the in-plane polarized closure domains of the (Pt/Co) multilayer and the NiFe magnetization. Similarly, exchange bias was observed in FeCoB in-plane magnetized layer, exchange coupled to out-of-plane magnetized $SmCo_5$ ferromagnet [16].

In recent years, the concept of topological protection of spin textures has received great interest, often involving closed shapes of magnetization vector fields, e.g. on spheroids in 3D or along loops in 2D. This is in particular the case of skyrmions, found in systems lacking inversion symmetry and displaying Dzyaloshinskii-Moriya interactions [17, 18]. These spin textures take the form of tubes in bulk systems [19-22] and bubbles in two-dimensional multilayered systems [23,24]. They were proposed to encode information at high density [25]. However, despite also displaying chiral features, the origin and magnetic stability of the 360° domain wall rings in in-plane magnetized exchange biased systems [25-32] derive from different physical mechanisms and have a different topology (Fig.1a). In this context of similarities yet with key differences, it is interesting to note that skyrmions have been stabilized at room temperature and zero applied magnetic field thanks to the use of exchange bias [33] (Fig.1b).

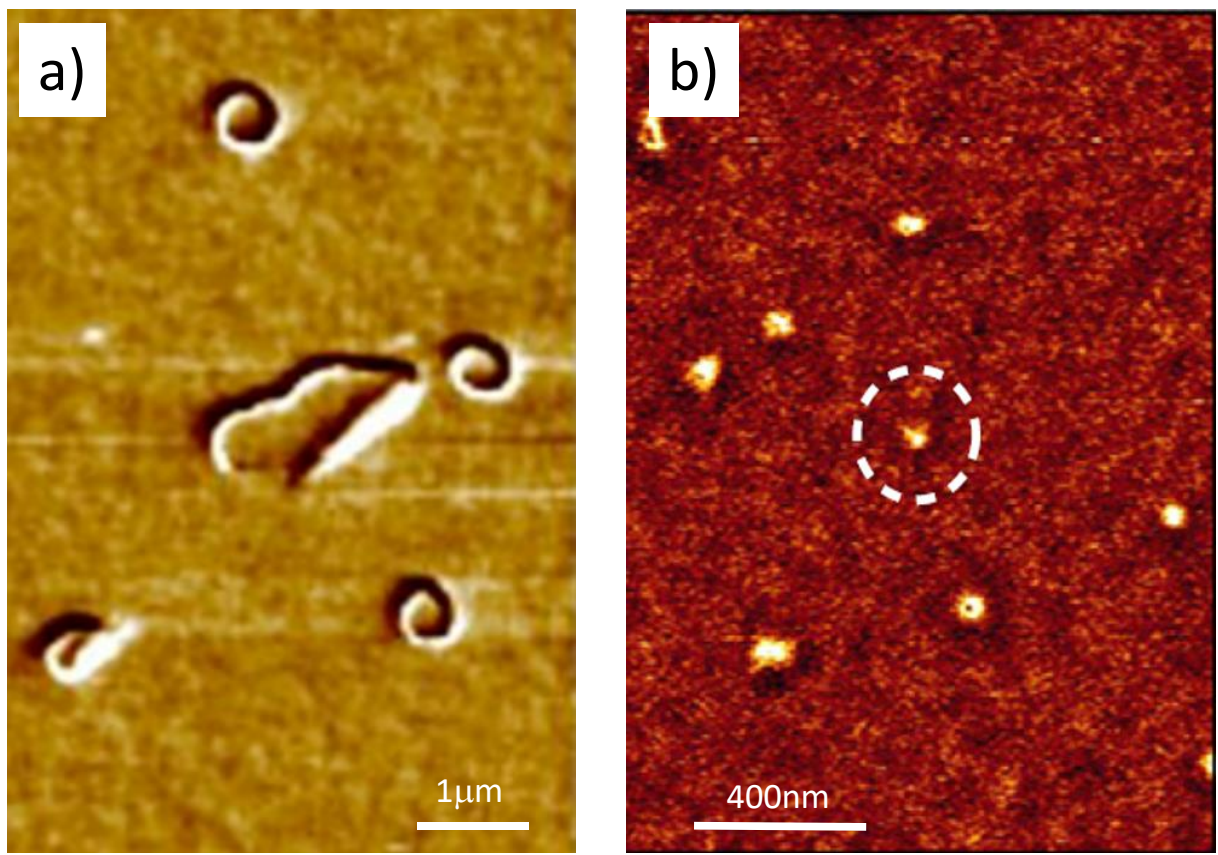

*Fig.1: MFM images of (a) 360° Néel-domain-wall rings in Co(3)/IrMn(7), displaying a chiral charge distribution; (b) skyrmions stabilized at room temperature and at zero field in Ta/Pt(3)/Co(0.3)/Ni80Fe20(2.1)/Ir20Mn80(4.2 )/Pt(3) (image adapted from Ref [33]).*

In this article, we discuss and compare spin textures formed in exchange biased magnetic structures with in-plane and out-of-plane magnetic anisotropy.

**2. 360° Domain Wall (360 DWs) segments and rings in exchange biased systems with in-plane magnetization:**

The formation of 360° domain walls in exchange-biased FM films have been a subject of interest over the last 30 years for both their practical implications in spintronic devices as well as to understand the fundamental physical mechanisms and material properties that control their formation, shape, size, and stability against temperature, applied magnetic fields and spin currents. Indeed, while 360° DWs in the form of segments and bubbles have been observed in ferromagnetic single films [34-36] and strips [37], they are much more common in exchange-biased films. In these systems, 360 DWs cause noise in spin-valve magnetic sensors as they constitute micromagnetic objects with strong irreversibility. Similarly, in MTJs, they are known to cause degradation in signal to noise ratio [18]. 360° DWs can be observed in the form of open loop segments (see for instance at the center of Fig.1a) or rings. In the case of segments, the two ends are pinned by defects, while the 360° DW rings are stabilized by their close loop shape combined with the easy-plane shape anisotropy. The rings are usually more stable than the segments. Once such ring structures are formed, they can remain stable at applied magnetic fields up to several tens of mT, hence hindering the operating specification range of sensor devices [25].

Dean *et al.* studied the formation mechanism of 360 DWs in an exchange-biased bilayer of CoFeB/IrMn [7]. They incorporated the measured magnetic properties of the bilayer into a micromagnetic model of exchange bias for granular AFM films and results were compared to imaging of the micro-magnetic structure determined from Lorentz TEM. The following are the salient conclusions of their study: i) to generate a stable 360° DW, one requires multiple nuclei of reversed domains having rotated all clockwise or anticlockwise, generating 180° Néel domain walls of same sense of rotation facing each other in the various nuclei, which under a sweeping magnetic field are pushed together to form a 360° DW. Application of the magnetic field at a small angle with respect to the EB direction promotes the formation of nuclei with the same sense of rotation; ii) Multiple nucleation regions emerge naturally as the exchange field acting on the FM is spatially non-uniform. This is a consequence of the distribution of anisotropy axis in the IrMn grains. In contrast, samples with minimal crystalline in-plane dispersion, such as epitaxial Fe (001)[100]//IrMn(001)[110], bilayers are less prone to form nucleation regions with reversed magnetization, and experiments confirm that no 360° DW are observed in such samples [38]; iii) the stability and density of 360° DW arises from the balance between intergrain exchange and anisotropy energies.

The internal spin-structure of the ring could be determined from Lorentz imaging as well as MFM imaging, as illustrated in Fig.2 [39]. The domain wall ring consists of a closed loop 360° Neel wall, i.e., magnetization rotates with a unique sense by 360° from the inner edge to the outer edge of the wall. However, at a given radial position in the ring, the spin orientation is uniform around the ring. This means, 360° domain walls located diametrically opposed in the ring have opposite chirality, contrary to the case of skyrmions. The magnetostatic charges generated by this magnetization configuration are represented in Fig.2 as well as the magnetic contrast expected from this charge distribution when imaging this spin texture with an MFM tip magnetized out-of-plane, which fits the experimental result shown in Fig.1a. The charge distribution is therefore chiral, which again is contrary to the case of skyrmions, whose charge distribution is invariant upon rotation.

Such 360° domain wall ring textures in in-plane exchange-biased films are quite stable, requiring fields of several tens of mT to be annihilated. However, they are not topologically protected due to the uniform magnetization for any given radial distance. Expressed in topological terms, the winding number [40] $S = \frac{1}{4\pi} \int n(x,y)\,\mathrm{d}x\mathrm{d}y$ with $n(x,y) = m \cdot (\partial_x m \times \partial_y m)$ is equal to zero since these spin textures are fully in-plane. This spin texture could in principle unwind through the 360° rotation of the central part. In practice, this does not occur as, at zero field, the core magnetization as well as the edge magnetization are maintained along the exchange bias field direction. However, this spin texture can also unwind by out-of-plane 180° flipping of the magnetization located at mid-width of the 360° ring. The energy barrier protecting these rings therefore arises from the in-plane confinement of the spins, and not from the creation of a Bloch point micromagnetic defect, as in the case of so-called topological protection.

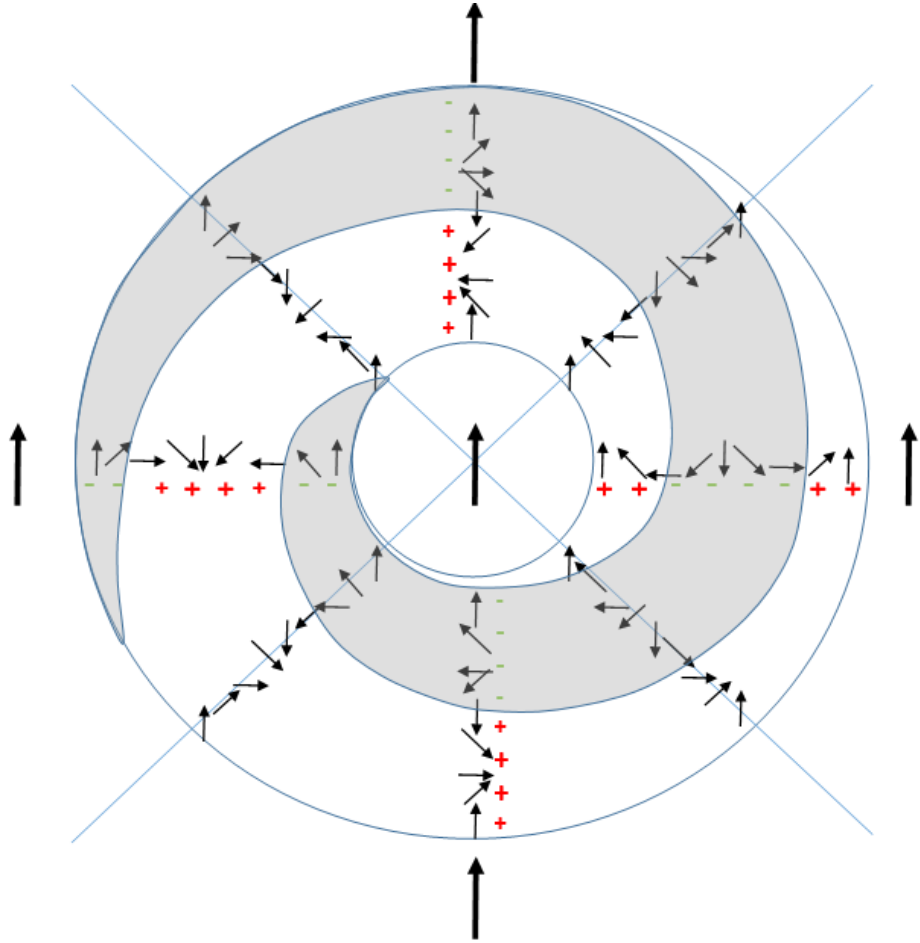

*Fig.2: magnetization and charge structure of 360° DW ring. The local density of magnetostatic charges determined by the divergence of magnetization is represented in red (positive charges) and green (negative charges). The white and grey contrasts indicate the expected MFM contrasts corresponding to those experimentally observed in Fig.1a.*

Note that in a given experimental realization in a sample, all rings have the same left hand/right hand dark/bright contrast (See for instance in Fig.1a, this is also evident in similar images reported in literature [7,27,29,31,32]). This implies that all 360° DW have the same sense of rotation from inside to outside the rings. This means that all (or at least a majority of) nuclei must have rotated along the same sense at the nucleation stage; otherwise there would be an equal share of right and left contrast spirals. Note that, contrary to the case of skyrmions in perpendicularly-magnetized films, the Dzyaloshinskii-Moriya interaction cannot explain this chirality in the charge distribution. Indeed, as the DMI vectors are oriented in the plane of the films for interfacial-based DMI, there is no DMI energy for purely in-plane-magnetized films.

As already mentioned, the selection of a given chirality of the rings is generally understood by considering their formation mechanism, as there always exists a slight misalignment between the exchange anisotropy field and the applied field, typically of a few degrees. The exchange anisotropy field is commonly set by the magnetic field applied during cooling the AFM/FM system from above the AFM blocking temperature [1]. Let's then assume that an hysteresis loop measurement is performed with the field applied slightly tilted by a few degrees away from the field cooling direction in the anticlockwise direction. Starting from positive saturation and decreasing the field, the magnetization will tend to rotate towards the exchange anisotropy field, i.e. in the clockwise direction. In the ideal situation of a perfectly uniform exchange field, all magnetization vectors would rotate along the same direction, leading to a coherent reversal or nucleation-propagation reversal. However, it is known that the exchange anisotropy field is inhomogeneous due to the distribution of the anisotropy axis in the AFM grains, on account for example of variations in crystalline growth orientation, and to the frustration of exchange interactions at the FM/AFM interface. This was emphasized by Malozemoff, who proposed a random field model of exchange anisotropy at a rough FM/AFM interface [41]. This random field adds up to the average exchange anisotropy field. Wherever the sum of this random field plus the exchange anisotropy field makes an opposite angle with respect to the applied field direction, a correlated area of magnetization forms whose moment rotates in the direction opposite to that of the majority of other moments, say, counterclockwise. As the field is progressively reversed, both the net magnetization and the correlated areas rotate away from their initial direction. Due to the misalignment of the applied field with the mean exchange anisotropy field, one of the two subsets of correlated areas forms a majority, while the other is a minority, thus progressively forming islands inside a matrix deriving from the former correlated areas. When the field gets close to negative saturation, each nucleated island is surrounded by a closed loop 360° Neel DW. These DWs have all the same left hand/right hand chirality, which is determined by the sign of the angular misalignment between the net exchange bias field direction and the applied field. The fact that most of these islands have the shape of circular rings of approximately the same diameter (see Fig.1a) may originate from an energy minimization, the shrinking being stopped at some point due the inability to unwind associated with the in-plane spin confinement associated with the easy-in plane shape anisotropy, or curvature-induced magnetostatic repulsion.

## 3. Skyrmions and skyrmoniums in exchange biased FM with perpendicular anisotropy:

Skyrmions have the same topology as a bubble with 180° wall in a perpendicularly-magnetized medium, with rotation symmetry. In thin films, the domain wall is of Néel type despite their dipolar energy, promoted by the Dzyaloshinskii-Moriya interaction (DMI). DMI makes skyrmions a chiral object, in which Néel walls point radially outwards or inwards depending on the sign of the DMI interaction, and whether the skyrmion has a center pointing up surrounded by down, or the reverse [44,45]. Skyrmions are characterized by the winding number +1, reflecting their chiral nature and the fact that magnetization spans $4\pi$ on the unit sphere over the entire spin texture.

Isolated skyrmions are metastable objects, which may be obtained following specific magnetization or current processes. A large DMI interaction is desirable to promote the stability of skyrmions, especially under dynamic conditions. Above a given threshold of DMI interactions, the ground state is stripe domains under zero magnetic field, a periodic network of skyrmions for moderate field, and isolated metastable skyrmions shortly before saturation. Skyrmions are therefore quite sensitive to an applied magnetic field, which influences their diameter through compression, similar to the 360° DW rings in in-plane EB systems. Exchange bias of FM with perpendicular anisotropy was recently utilized [33,42] to stabilize skyrmions at room temperature and zero applied external magnetic fields (Fig.1b). The interlayer exchange coupling acts as an effective magnetic field that breaks the symmetry of "up" and "down" domains, resulting in the formation of skyrmions inside one single domain with opposite magnetization. In this study, IrMn was used as the AF, which is a very common choice. Ma *et al*. calculated and observed that Dzyaloshinskii-Moriya Interactions (DMI) exist across FM/AFM interfaces of IrMn/CoFeB/MgO multilayer thin films [43], with a DMI vector in the plane of the interface. This extra DMI adds up to the usual one of skyrmionic films promoted by other interfaces, and contributes to impose a uniform chirality of the domain walls as in standard skyrmions. In particular, note that the contrast is rotationally-invariant, contrary to 360° domain-wall rings for in-plane magnetized exchange-biased systems (fig.2).

Skyrmionic textures with 360° domain walls also exist, called skyrmoniums (see fig.3). These are again metastable ring textures, now spanning $8\pi$ over the unit sphere for magnetization and winding number 2, which can be promoted by specific magnetizing or current history [46]. Their distribution of dipolar charges is rotational-invariant, similar to skyrmions, and therefore distinct from the in-plane 360° DW rings in EB systems.

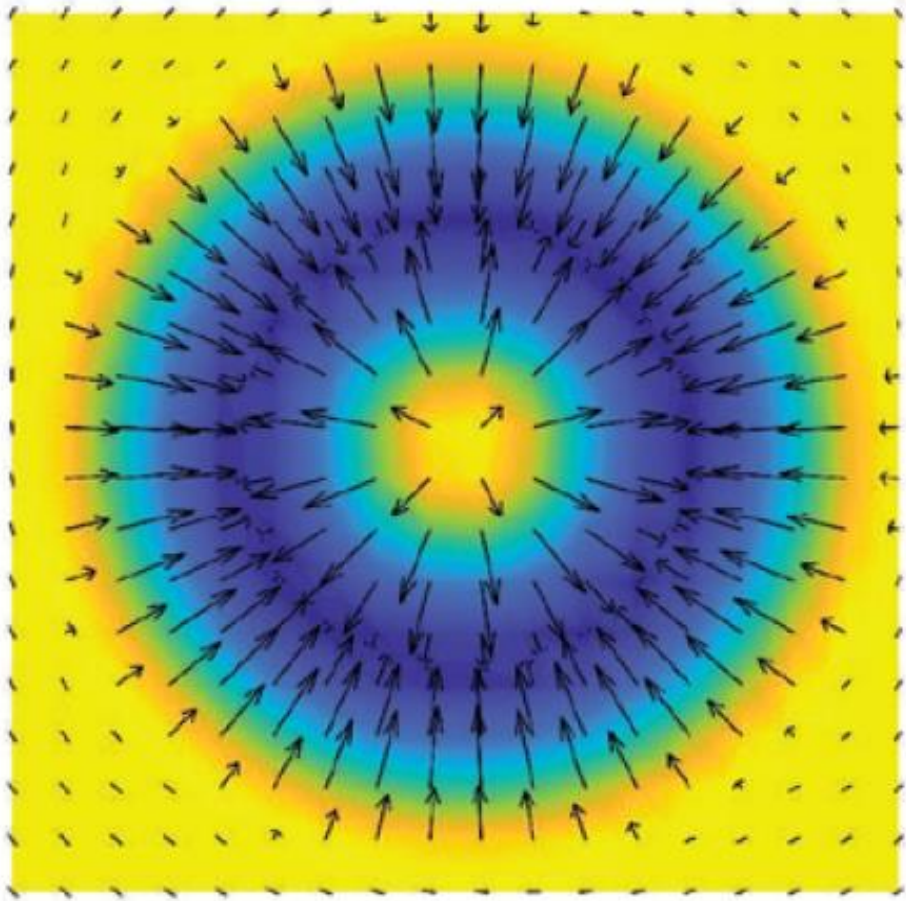

Fig.3: Skyrmonium structure. From [46]

**Conclusion:**

We have compared spin textures that develop in exchange biased systems with either in-plane or out-of-plane magnetic anisotropy. For in-plane exchange biased systems, 360° Neel domain wall rings may form, which are quite stable when subjected to an external magnetic field. Their stability originates from the closed loop shape of the domain wall combined with the in-plane confinement of the spins due to the easy-plane shape anisotropy. Their spin texture is not chiral and has zero winding number, but the distribution of associated dipole charges is chiral, reflecting the direction of rotation of the in-plane magnetization during the initial stages of the reversal. Out-of-plane exchanged-biased systems may be influenced by Dzyaloshinskii-Moriya Interactions, whereas DMI are absent in systems with in-plane anisotropy. As a result, DMI promotes the formation of skyrmions and skyrmoniums, with 180° and 360° DWs and winding number 1 and 2, respectively, similar to thin films with no EB. In contrast to in-plane magnetized systems, their distribution of dipolar charges is not chiral.

**Acknowledgements**: E.M. acknowledges financial support from CNRS for a sabbatical at SPINTEC. We acknowledge fruitful discussions with our colleague O. Boulle.

**Conflict of interest**: The authors declare that they have no conflict of interest in relation to this research.

**Data availability**: The authors make the data supporting this paper available upon reasonable request.